\documentclass[modern]{aastex61}

\usepackage{graphicx}
\usepackage{amsmath}
\usepackage{amssymb}
\usepackage{enumitem}

\newcommand\mybar{\kern1pt\rule[-\dp\strutbox]{.8pt}{\baselineskip}\kern1pt}

\setlist[itemize]{noitemsep, topsep=0pt, leftmargin=*}

\shorttitle{Titan and the Early Universe}
\shortauthors{Loeb}

\begin{document}

\title{Life on Titan May Signal Early Life in the Universe}

\author{Abraham Loeb}
\affiliation{Astronomy Department, Harvard University, 60 Garden
  St., Cambridge, MA 02138, USA}

\begin{abstract}
The temperature of the cosmic microwave background (CMB) was equal to
the surface temperature of Saturn's moon Titan, $94$K, at a redshift
$z=33.5$, after the first galaxies formed. Titan-like objects would
have maintained this surface temperature for tens of Myr irrespective
of their distance from a star.  Titan has the potential for the
chemistry of familiar life in its subsurface water ocean, as well new
forms of life in the rivers, lakes and seas of liquid methane and
ethane on its surface. The potential future discovery of life on Titan
would open the possibility that the earliest lifeforms emerged in
metal-rich environments of the earliest galaxies in the universe,
merely 100 Myr after the Big Bang.
\end{abstract}

\section{Introduction}

The cosmic microwave background (CMB) provides a universal heating
source of temperature~\citep{1996ApJ...473..576F}, $T_{\rm cmb}=94 {\rm K} \times
[(1+z)/34.5]$ at a cosmological redshift $z$. Interestingly, this
temperature matches the surface temperature of Saturn's largest moon,
Titan at $z\sim 34$, about 90 Myr after the Big Bang. Hence, a
Titan-like object at that early time would have maintained this
temperature for tens of Myr, sufficient for primitive life to form in
its liquid reservoirs or atmosphere, irrespective of its distance from
a star.

\section{First Objects}

The standard cosmological model predicts that the first generation of
stars and galaxies formed before $z\sim
34$~\citep{2013fgu..book.....L}. Based on the measured cosmological
parameters \citep{2020A&A...641A...6P}, the first star-forming halos
collapsed at $z\sim 71$ on our past light cone and at $z\sim 77$
within the entire Hubble volume~\citep{2014IJAsB..13..337L}, including
the delay from the streaming motion of baryons relative to dark
matter~\citep{2012MNRAS.424.1335F}.

Hydrodynamical cosmological simulations predict that the first
galaxies formed population III stars that were predominantly massive
\citep{2013fgu..book.....L}. For massive stars that are dominated by
radiation pressure and shine near their Eddington luminosity $L_{\rm
  E}=1.3\times 10^{39}~{\rm erg~s^{-1}}(M_\star/10M_\odot)$, the
lifetime is independent of stellar mass $M_\star$ and is universally a
few Myr, set by the nuclear efficiency of converting rest-mass to
radiation, 0.7$\%$, namely $\sim (0.007M_\star c^2)/L_{\rm E}= 3~{\rm
  Myr}$~\citep{2001ApJ...552..464B}.

Consequently, the subsequent delay in dispersing heavy elements from
the first stellar winds or pair-instability supernovae could have been
as short as a few Myr, only a few percent of the age of the Universe
at $z\sim 34$. The supernova ejecta could have produced
high-metallicity islands that were not fully mixed with the
surrounding primordial gas, leading to efficient formation of planets
and moons within them.

Altogether, this suggests that massive stars and supernovae were able
to enrich the interstellar medium in the cores of the earliest
galaxies with heavy elements before $z\sim 34$, leading to metal-rich
pockets of gas inside of which the second generation of stars could
have formed, accompanied by Titan-like objects.

\section{Prospects of Life on Titan}

The temperature coincidence between Titan's surface and the CMB at
$z\sim 34$ raises the fascinating possibility of testing how early
life could have arisen in the Universe by studying Titan. In other
words, the question of whether Titan hosts life has cosmic
implications.

In the Solar system, Titan is the only object besides Earth that has
rivers, lakes and seas on its surface, as well as a cycle of methane
and ethane liquids raining from clouds, flowing across its surface and
evaporating back into the atmosphere, similarly to Earth's water
cycle. Titan is also thought to have a subsurface ocean of water. Its
atmosphere is primarily nitrogen like Earth's, but with a $\sim 5\%$
contribution of methane. Titan's landscape is covered with dark dunes
of hydrocarbon grains, primarily around the equatorial regions.

Gravity measurements by the Cassini spacecraft revealed that Titan has
an underground ocean of liquid water, likely mixed with salts and
ammonia~\citep{2019SSRv..215...33L}. Radio signals detected by the
Huygens probe in 2005 strongly suggested the presence of an ocean
55-80 km below the icy surface, allowing for the chemistry of
life-as-we-know-it. In addition, Titan's bodies of liquid methane and
ethane might serve as a foundation for the chemistry of
life-as-we-do-not-know-it on the moon's surface.

Whether the physical conditions on Titan gave birth to these forms of
life is unknown.  The realization that Titan's atmosphere is rich in
organic compounds led to the proposal that it produced the chemical
precursors of life. In particular, stable cryogenic cell membranes
could arise from compounds observed in Titan's
atmosphere~\citep{2015SciA....1E0067S}. The proposed chemical base for
these membranes is acrylonitrile, which was detected in Titan's
atmosphere by Cassini and
ALMA~\citep{2017ApJ...844L..18D,2017SciA....3E0022P}. Analysis of data
from the Cassini-Huygens mission reported anomalies in the atmosphere
near the surface which could be consistent with the presence of
lifeform of methane-consuming organisms, but may alternatively be due
to abiotic chemical
processes~\citep{2010Icar..208..878S,2010JGRE..11510005C}.

Laboratory experiments~\citep{2010DPS....42.3620H} indicate that when
discharge power is applied to a combination of gases like those in
Titan's atmosphere, they make prebiotic molecules such as the five
nucleotide bases of DNA and RNA as well as amino-acids, among many
other compounds.

\section{Discussion}

The thermal gradients needed for life can be supplied by geological
variations on the surface of early Titan-like objects. Examples for
sources of free energy are geothermal energy powered by the object's
gravitational binding energy at formation and radioactive energy from
unstable elements produced by the earliest supernova.  If life
persisted at $z\lesssim 34$, it could have also been transported to
newly formed objects through panspermia~\citep{2018ApJ...868L..12G}.

Given the above considerations, the search for life on Titan could
open the possibility that life may have started at a redshift of
$z\sim 34$ in the standard cosmological model of our Universe.

In addition to studying Titan, the feasibility of life in the early
universe can be further tested by searching for planets with
atmospheric bio-signatures around low-metallicity stars in the Milky
Way galaxy or its dwarf galaxy satellites. Such stars represent the
closest analogs to the first generation of stars at early cosmic
times.
 
\bigskip
\bigskip
\section*{Acknowledgements}

This work was supported in part by Harvard's {\it Black Hole
  Initiative}, which is funded by grants from JFT and GBMF.  

\bigskip
\bigskip
\bigskip

\bibliographystyle{aasjournal}
\bibliography{t}
\label{lastpage}
\end{document}